\documentclass[letterpaper, 10 pt, conference]{ieeeconf}  

\IEEEoverridecommandlockouts                              

\usepackage{graphics} 
\usepackage{epsfig} 
\usepackage{float} 

\usepackage{mathptmx} 
\usepackage{times} 
\usepackage{amsmath} 
\usepackage{amssymb}  

\usepackage{dsfont}
\usepackage{soul}
\usepackage[normalem]{ulem}
\usepackage[utf8]{inputenc}

\usepackage{graphicx}

\usepackage{amsmath}
\usepackage{booktabs}
\usepackage{algorithm}
\usepackage[noend]{algorithmic}
\usepackage{amsfonts}
\usepackage{blindtext}
\graphicspath{{Figures//}}
\DeclareGraphicsExtensions{.eps,.pdf,.png,.jpg,.tif,.tiff,.ps}

\usepackage{tabu}

\usepackage{outlines}

\usepackage{breqn}

\makeatletter
\let\NAT@parse\undefined
\makeatother
\usepackage[hidelinks]{hyperref}  
\usepackage{url}
\usepackage[nameinlink,capitalize]{cleveref}
\usepackage{soul}
\usepackage[colorinlistoftodos]{todonotes}

\definecolor{navy}{rgb}{0.1, 0.1, 0.8}
\definecolor{gray}{rgb}{0.6, 0.6, 0.6}
\definecolor{myblue}{rgb}{.8, .8, 1}
\definecolor{olive}{rgb}{0.1, 0.5, 0.1}
\definecolor{magenta}{rgb}{0.55, 0.0, 0.55}
\usepackage{adjustbox}
\usepackage{textcomp}
\usepackage{siunitx}

\usepackage{lipsum}

\usepackage{xspace}

\newcommand{\citet}[1]{\citeauthor{#1}~\shortcite{#1}}
\newcommand{\citep}{\cite}

\usepackage[belowskip=-15pt,aboveskip=0pt]{caption}
\usepackage{subcaption}

\newcommand{\titlename}{Traffic disruption modelling with mode shift in multi-modal networks}

\title{\LARGE \bf \titlename}

\author{Dong Zhao$^{1}$, Adriana-Simona~Mih\u{a}i\c{t}\u{a}$^{1}$, Yuming Ou$^{1}$, 
Sajjad Shafiei$^{2}$, Hanna Grzybowska$^{3}$, \\
Kai Qin$^{2}$, Gary Tan$^{4}$, Mo Li$^{5}$ and Hussein Dia$^{2}$
\thanks{$^{1}$ University of Technology Sydney, Ultimo, NSW 2007, Australia. Corresponding authors contact: {\tt\small Dong.Zhao@student.uts.edu.au}}
\thanks{$^{2}$ Swinburne University of Technology, Hawthorn, VIC 3122, Australia} 
\thanks{$^{3}$ Data61, CSIRO, Eveleigh, NSW 2015, Australia}
\thanks{$^{4}$ National University of Singapore, Singapore}
\thanks{$^{5}$ Nanyang Technological University (NTU), Singapore}
}

\begin{document}

\maketitle
\thispagestyle{empty}
\pagestyle{empty}

\begin{abstract}
A multi-modal transport system is acknowledged to have robust failure tolerance and can effectively relieve urban congestion issues. However, estimating the impact of disruptions across multi-transport modes is a challenging problem due to a dis-aggregated modelling approach applied to only individual modes at a time. To fill this gap, this paper proposes a new integrated modelling framework for a multi-modal traffic state estimation and evaluation of the disruption impact across all modes under various traffic conditions. First, we propose an iterative trip assignment model to elucidate the association between travel demand and travel behaviour, including a multi-modal origin-to-destination estimation for private and public transport. Secondly, we provide a practical multi-modal travel demand re-adjustment that takes the mode shift of the affected travellers into consideration. The pros and cons of the mode shift strategy are showcased via several scenario-based transport simulating experiments. The results show that a well-balanced mode shift with flexible routing and early announcements of detours so that travellers can plan ahead can significantly benefit all travellers by a delay time reduction of 46\%, while a stable route assignment maintains a higher average traffic flow and the inactive mode-route choice help relief density under the traffic disruptions.
\end{abstract}


\section{Introduction} 
\label{I_Introduction}
\subsection{Background and motivation} \label{I_A_Background_and_motivation}
Resilient cities have recently embraced a fully-connected multi-modal transport network that gives travellers more freedom when choosing when, where and how to travel. However, multi-modal urban environments are also vulnerable due to the lack of tolerance against an ever-growing population, an expanding travel demand, a high private car ownership, deficient transport design, inadequate traffic control and flawed travelling or driving behaviour \cite{Rahman2021}. 


To improve the efficiency of the transport system at a large scale, the encouragement of a travel behaviour change and active mode shift is an encouraging option studied recently \cite{Ettema2016}. Many other research studies reinforce this initiative by providing substantial evidence via data-driven, or simulation-based approaches \cite{Wen2018, Mihaita2018, Mao2021}. The data-driven approaches capture the real traffic behaviour before and after disruptions, and some applications are used in programs such as: INPHORMM, TAPESTRY or Travel Smart \cite{Ma2017}. Other early studies revealed the value of public transport by investigating the change of traffic states (e.g. section flows, traffic volumes or travel times) and proposed an entire public transport service removal when massive public transport disruptions occur or when service is suspended \cite{Adler2016, Moylan2016}. Few studies that consider a simulation approach mention that the change in the level of congestion before and after the removal of public transport services would clarify the significance of public transport \cite{NguyenPhuoc2018}. More recently, the unprecedented COVID-19 pandemic has heavily modified the travel demand and provided evidence with regards to the impact of traffic demand across all mode shifts in a city \cite{Das2021}. 

\textbf{Challenges:} 
All previous studies solve the mode choice problem before departing, and most publications provide modelling methods from a macroscopic or a mesoscopic level based on a statical analysis. There is little research into investigating the benefits of an active mode shift from a dynamic microscopic perspective and its impact when traffic disruptions occur. A significant gap is present due to the lack of data regarding the impacted demand under incidents and active mode shifts. Some studies rely on surveys or a stated preference obtained ahead of trips to obtain the number of impacted travellers or the number of mode and route shift \cite{Auld2020, Guzman2021}. However, we emphasise identifying the impacted origin-to-destination (OD) trips affected by disruptions in a simulating model, and the change of mode and route choice that leads to a demand change is employed for evaluating the impact on network performance in our work. 

Apart from the lack of data, quantifying the impact of disruptions is also a major challenge; some research studies have analysed the change of trip-based mean delay, mean speed \cite{Aftabuzzaman2010} or travel time \cite{NguyenPhuoc2018}. However, such indicators can hardly differentiate the impact from the general traffic (e.g. recurrent congestion) or from traffic control strategies. To address this issue, we work across several indicators versus baseline conditions in order to evaluate the efficiency of the proposed ones. 
\begin{figure*}[!htb]
    \centering
    \includegraphics[width=0.95\textwidth, height = 8cm]{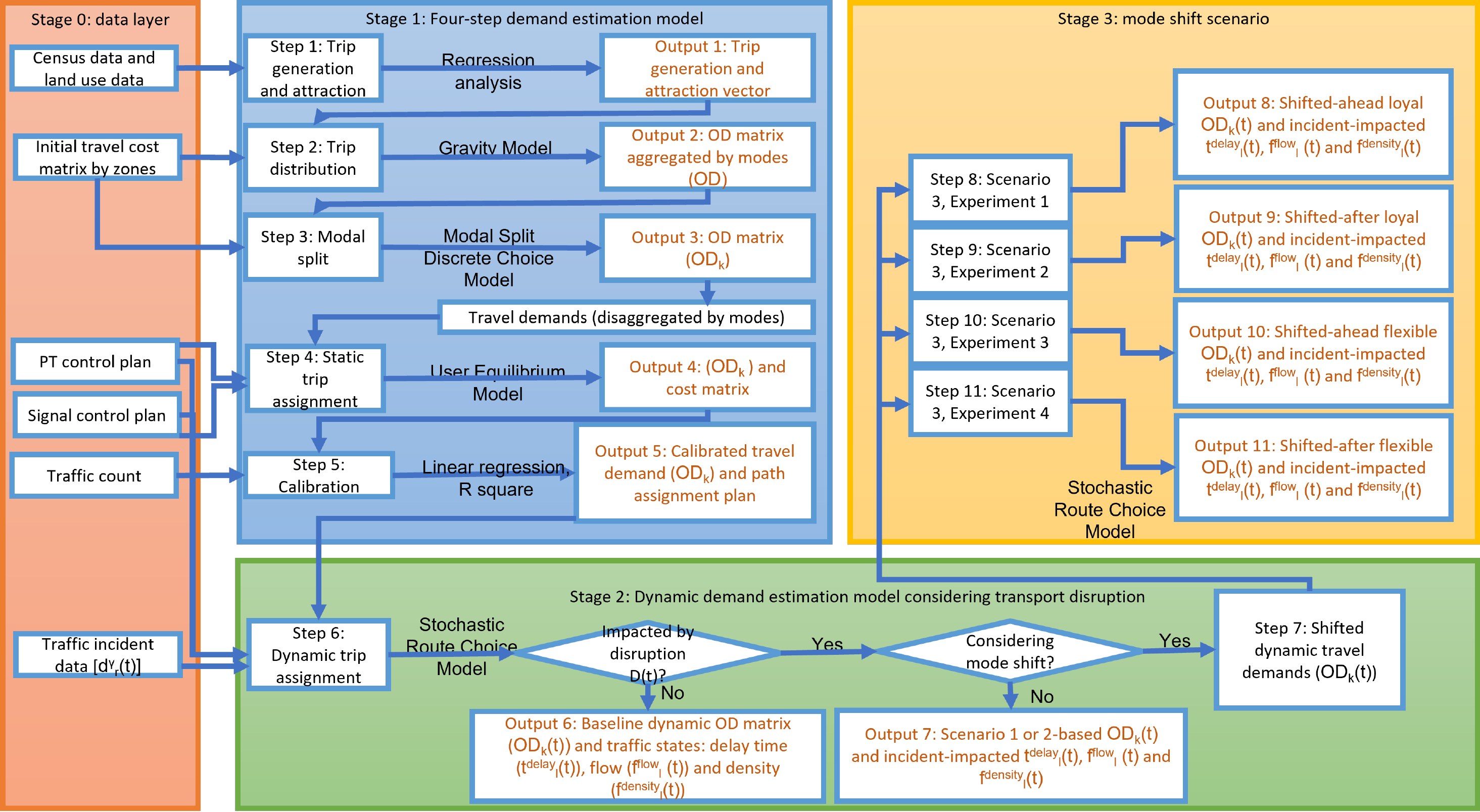}
    \caption{\footnotesize Framework of our proposed multi-modal transport network modelling under disruptions.}
    \label{Fig_research_framework}
\end{figure*}

Another major challenge of dynamically simulating the mode shift is the lack of dynamic demand data and the method of integrating the OD estimation across different transport modes in order to identify the impacted trips. Most previous research studies only consider a single-mode \cite{Thompson2019}, while some research studies model car-based transportation versus public transportation differently \cite{Hussain2021}. Extensive evidence considers the OD estimation from a total generation and attraction data based on the gravity model. This method has been largely developed with the improvement of mathematical, analytical and computational skills. However, the large potential of the gravity model approach in the transport field has not yet been fully explored, as most research studies attempted to investigate the OD matrix for a single transport mode, mostly cars. There is still a need to consider the influence of other transport modes when mode splitting and trip assigning under a multi-modal public transport environment. The challenge of integrating the OD matrices of various public transport modes with that of private vehicles is still unsolved.

\subsection{Paper Contributions}\label{I_B_Contributions}
In this paper, we propose an integrated modelling approach comprised of multiple stages, from the data selection and filtering to the origin-to-destination estimation modelling across multiple modes, down to a dynamic assignment and microscopic simulation modelling aimed at evaluating the impact of disruptions across multiple modes. Finally, we propose a mode shift impact modelling to evaluate the best mitigation strategies and employ different disruption impact indicators such as delay time, flow, density and travel time for identifying the impacts. 

Another important contribution represents the investigation of the mode shift behaviour according to dynamic traffic states; more specifically, we provide a method to examine the change in traffic states and the travel costs due to mode and route shifts under traffic disruptions. We model the decision-making en-route and the mode choice relies on an iterative traffic assignment; this means that, for those flexible travellers, the route choice is modifiable during their travelling, and the decision-making is more appealing than those travellers who are loyal to the initial routing plan. 
To summarise, the main theoretical and methodological contributions of this paper are: 
\begin{itemize} 
\item an integrated OD estimation modelling framework for multi-modal transport networks, 
\item a suitable spatial-temporal disruption impact modelling via a multi-modal transport simulation approach, 
\item a dynamic traffic assignment model that simulates the mode shift behaviour via a dynamic demand adjustment, 
\item an analytic method regarding the mechanism of mode shift as well as its impact under traffic disruptions.
\end{itemize}

This paper is organised as follows. In \cref{II_methodology}, the dynamic trip assignment model is discussed, and the details of the integrated OD estimation is highlighted in \cref{II_B_Integrated_OD_estimation}, followed by the methodology of traffic disruption modelling and the procedures for determining the spatial-temporal impact of traffic disruptions in \cref{II_C_Disruption_modelling_disruption_impact_identification}. The model for mode shift and impacted trips identification are included in \cref{II_D_Mode_route_shift_modelling}. The application of the proposed methods to a real network is presented in \cref{III_Case_study} and the results of the case study are demonstrated in \cref{IV_Results}. Finally, the research conclusion and the future directions are provided in \cref{V_Conclusion}. 

\section{Methodology}\label{II_methodology}
\subsection{Modelling framework}\label{II_A_Research_framework}
\cref{Fig_research_framework} showcases our proposed modelling framework for evaluating the impact of disruptions across multi-mode transport networks. The framework consists of three stages: at \textit{Stage 0} we collect, filter and aggregate all the input data-sets (such as traffic flow counts, traffic control plans, incident logs, etc.); at \textit{Stage 1} we propose a multi-modal demand estimation modelling with the purpose of obtaining an integrated multi-modal OD demand matrix (see details in \cref{II_B_Integrated_OD_estimation}); at \textit{Stage 2} we further propose a dynamic trip assignment and demand refinement based on the impact of transport disruptions (as explained in \cref{II_C_Disruption_modelling_disruption_impact_identification}), and finally, at \textit{Stage 3} we construct various mode and route shift strategies and their impact on the traffic congestion, as further described in \cref{II_D_Mode_route_shift_modelling}. 

\subsection{Integrated multi-modal OD estimation \textit{(Stage 1)}} \label{II_B_Integrated_OD_estimation}
The multi-modal transport system is firstly modelled by implementing a four-step demand estimation model but adapted to multiple public transport modes, as shown in \textit{Stage 1}- \cref{Fig_research_framework}, including trip generation and attraction \textit{(Step 1)}, trip distribution \textit{(Step 2)}, modal split \textit{(Step 3)} and static trip assignment at the macroscopic level \textit{(Step 4)}, followed by a calibrated travel demand path assignment plan \textit{(Step 5)}. The study area consists of $Z_j,\ j=\{1\ldots J\}$ zones, and for each time period $t$, the travel demand matrix, which is the main output of this stage \textit{(Output 5)}, is denoted as:
\begin{equation} \label{equation_II_B_1}
    {OD}_k(t)=\left[T_{i,j}^k(t)\right]_{Z\times Z},\ i,j=\{1\ldots J\}, 
\end{equation}
where $T_{i,j}^k$ stands for the number of trips originating from zone $i$ and arriving at zone $j$ at time interval $t$ by transport mode $k$. Due to space constraints, we provide the mathematical modelling of the Gravity-Model for multi-mode public transport in the supplementary material \cite{appendix}, while focusing in the following on the incident impact modelling. 
\subsection{Disruption modelling without mode shift (\textit{Stage 2})}\label{II_C_Disruption_modelling_disruption_impact_identification}
By using the calibrated OD demand from \textit{Stage 1} and the reported incident logs, we further apply a dynamic trip assignment which re-adjusts and generates a dynamic and time-dependent OD matrix that is used for: a) evaluating a baseline scenario where people travel as usual, without any disruptions (see \textit{Output 6}), and b) evaluating the impact of reported accident logs but assuming that people do not make any changes to their trips, and instead wait for the incident to be cleared off (see \textit{Output 7}).

\subsubsection{Indicators for impact identification}\label{II_C_1_impact_identification}
The indicator we use when determining the impact of a traffic disruption $D$ on link $l$ during time period $t$ is the ratio of the traffic state parameter $v$, which is illustrated by the following formulas:
    \begin{equation} \label{equation_II_C_1}
        R_l^D (t)=\frac{\left(v_l^\alpha - v_l^{D,\beta}\right)\left(t\right)}{v_l^\alpha\left(t\right)}
    \end{equation}
    \begin{equation} \label{equation_II_C_2}
        \Delta v_l^D (t)=\left(v_l^\alpha - v_l^{D,\beta}\right)(t)
    \end{equation}   
where $v^{\alpha}$ is the traffic state not affected by disruptions, $v_l^{D,\beta}$ is the traffic state affected by the disruption and $\Delta v_l^D (t)$ represents the scale of the impact; finally $R_l^D (t)$ stands for the ratio of a baseline versus a disrupted network. The total number of links in the study area is $l, l=\{1\ldots L\}$; $\alpha$ represents the baseline traffic situation; $\beta$ represents the situation when the traffic disruption $D$ is impacting the network at time period $t$. 

To understand the change of traffic states, various indicators such as mean link delay time $t_l^{delay} (t)$, mean travel time $t_l^{travel} (t)$, flow $f_l^{flow} (t)$ or density $f_l^{density} (t)$ during time period $t$ are used for ratio impact analysis provided later in the Results section.

The overall traffic disruption in the network that impacts the traffic states is represented by:
    \begin{equation} \label{equation_II_C_3} 
        D(t)=g\left( \sum_{r=1}^{L}{d_r^{\gamma}(t)} \right)
    \end{equation}
where the impact of the disruption is the function of the sum of all appeared disruptions in the network during a time period $t$; $d$ is the individual disruption event; $r$ indicates the location of the disruption, $r=\{1\dots L\}$; $\gamma$ is the binary parameter that indicates whether the disruption is impacting or not the link $l$ during $t$.  

\textbf{Temporal impact identification:}
The total impact duration ($\Delta t_D^{impact}$) of the network is the accumulation of the time period from the time when the travelling of a vehicle is first impacted by the disruption $D$ ($t_D^{impact-initial}$) until the time when the first vehicle can travel as usual, without being impact, denoted as $t_D^{impact-end}$. Therefore, the total impact duration can be described by:
    \begin{equation} \label{equation_II_C_4}
        \Delta t_D^{impact}=t_D^{impact-end} - t_D^{impact-initial}
    \end{equation}
with additional constraints with regards to the initial time of the disruption ($t_D^{D-initial}$), the end of the disruption ($t_D^{D-end}$) and the parameters of traffic states ($v_l^{D,\beta}$ and $v_l^{\alpha}$):
    \begin{equation} \label{equation_II_C_5}
        t_D^{impact-initial} \geq t_D^{D-initial}
    \end{equation}
    \begin{equation} \label{equation_II_C_6}
        \epsilon \sum_{l=1}^{L}{v_l^{D,\beta}\left( t^{impact-initial} \right)} \geq \sum_{r=1}^{L}{v_l^{D,\alpha}\left( t^{impact-initial} \right)}
    \end{equation}
    \begin{equation} \label{equation_II_C_7}
        t_D^{impact-end} \geq t_D^{D-end}
    \end{equation}
    \begin{equation} \label{equation_II_C_8}
        \epsilon \sum_{r=1}^{L}{v_l^{D,\beta}\left( t^{impact-end} \right)} \leq \epsilon \sum_{r=1}^{L}{v_l^{D,\alpha} \left( t^{impact-end} \right)}
    \end{equation}
where $\epsilon$ is the factor of impacted traffic state that indicates the pre-defined level of impact. For instance, if $\epsilon$ is 90\% and the indicator is delay time, the link is assumed to be impacted by the disruption when 90\% of the link delay time is greater than the delay time for the baseline scenario. As for the representations of traffic states, such as travel speed, that are reduced by the disruption, the associated constraints are switch from \cref{equation_II_C_6} to:
    \begin{equation} \label{equation_II_C_9}
        \sum_{r=1}^{L}{v_l^{D,\beta} \left( t^{impact-initial} \right)} \leq \epsilon \sum_{r=1}^{L}{v_l^{D,\alpha} \left( t^{impact-initial} \right)}
    \end{equation}
from \cref{equation_II_C_8} to:
    \begin{equation} \label{equation_II_C_10}
        \epsilon \sum_{r=1}^{L}{v_l^{D,\beta} \left( t^{impact-initial} \right)} \geq \sum_{r=1}^{L}{v_l^{D,\alpha} \left( t^{impact-initial} \right)}
    \end{equation}
\textbf{Spatial impact identification via links:}
The time impact of the disruption is captured from a network-wide perspective, but the spatial impact can be analysed via a link-based analysis:
    \begin{equation} \label{equation_II_C_11}
        S^D(t)=[l^{\gamma} (t)]
    \end{equation}
where $\gamma$ is the binary parameter that indicates whether the disruption is impacting the link $l$ during $t$; similarly to the time impact, we define the following constraints when the impacted links can be identified if:
    \begin{equation} \label{equation_II_C_12}
        \epsilon v_l^{D,\beta}\left( t^{impact-initial} \right) \geq v_l^{D,\alpha}\left( t^{impact-initial} \right)
    \end{equation}
    \begin{equation} \label{equation_II_C_13}
        \epsilon v_l^{D,\beta}\left( t^{impact-end} \right) \leq \epsilon v_l^{D,\alpha} \left( t^{impact-end} \right)
    \end{equation}
The impact on the incident duration can be defined by using the affected links as follows:
    \begin{equation} \label{equation_II_C_14}
        \Delta t_{l,D}^{impact}=t_{l,D}^{impact-end} - t_{l,D}^{impact-initial}
    \end{equation}
\textbf{Spatial impact identification via OD matrix:}
In terms of the zonal impact analysis, the affected number of trips from an OD pair can be determined by comparing the OD matrix with or without the disruption in the road network. The level of impact on trips is subjected to the ratio of the disparity of trips travelled between zones in a baseline versus incident scenario, which is derived from \cref{equation_II_C_1} as follows: 
    \begin{equation} \label{equation_II_C_15}
        R_{i,j}^T(t)=\frac{\left(T_{i,j}^\alpha - T_{i,j}^\beta\right) \left(t\right))}{\left( T_{i,j}^\alpha + \epsilon \right)\left(t\right)}, T_{i,j}^\alpha, T_{i,j}^\beta and \epsilon \geq 0
    \end{equation}
where $T_{i,j}^\alpha$ indicates the number of trips between zone $i$ and $j$ under a baseline scenario, $T_{i,j}^{D,\beta}$ represents the disruption-impacted number of trips and $\epsilon$ is a small constant that is added at the number of trips without impacting by any disruption in order to enable calculation, as in reality, some of the zone pairs are inaccessible for a specific transport mode. The different between the baseline and the disruption-impacted trips is denoted as $\Delta T_{i,j}(t)$. Therefore, if the ratio of disparity trip is between 0 and 1, then the number of trips inside the affected OD is reduced; while if the ratio is negative, the network experiences an increase in the affected number of trips which can be explained by alternative modes being deployed in the network by the affected number of people. 

\subsubsection{Disruption modelling}
\label{II_C_2_disruption_modelling}
The impacted travel cost of every link during time $t$ due to the reduction of link capacity should be described as a function of disruption duration (\cref{equation_II_C_14}) and disruption scale (\cref{equation_II_C_11}):
    \begin{equation} \label{equation_II_C_16}
        C_l(t)=h\left(t,l \right), l(t) \in S^D(t), t \in \Delta t_D^{impact}
    \end{equation}
This means that, for each impacted link during the incident, the properties of the link can be highlighted by the capacity or by the limited travel speed change by time and location. Therefore, if we assume that the travel cost is majorly subjected to the travel speed, the link closure can be described as $C_l(t)=0$, where the limited travel speed equals zero during the disruption period; in terms of the simulation of link capacity reduction, this can be achieved by introducing a weight to the designed limited travel speed $V$ for each lane $w$ of the link $l$, which is denoted by $\eta_{l,w}$. Therefore, the weighted limited travel speed can be illustrated by $\eta_{l,w} V_{l,w}$. 

\subsection{Disruption modelling under mode shift (\textit{Stage 3})}\label{II_D_Mode_route_shift_modelling}
\subsubsection{Mode shift modelling}\label{II_D_1_Mode_route_shift_modelling}
The commonly used strategies to minimise travel costs are mode and route shifts. By mode shift, we refer to the shift of the travel demand between different means of transportation. Therefore, the objective of mode shift modelling is to first find the changeable demand for each transport mode, then adjust it accordingly in the travel demand matrix in order to evaluate its impact. Such process is what is illustrated as \textit{Step 7} and \textit{Step 8} with the highlighted \textit{Outcome 8} in \cref{Fig_research_framework}. 

Given a scenario of a road network disruption, the perceived travel cost increases, and the decision-making on mode and route choice is re-decided according to the utility function. The re-calculated set of the shortest path by a transport mode will update the travel demand. For the updated travel demand matrix, the number of impacted trips travelled by mode $k_1$ due to the disruption is transferred to demand travelled by mode $k_2$, the number of transferred trips equals $\delta T_{i,j}^{k_1,k_2}(t)$. Hence, the travel demand by a transport mode $k_1$ between each pair of zones is calibrated by the impacted demand:
    \begin{equation} \label{equation_II_D_1} 
        T_{i,j}^{k_1}(t)=T_{i,j}^{k_1}(t) - \delta T_{i,j}^{k_1,k_2}(t)
    \end{equation}
With regards to a mode $k_2$, the number of zonal trips is increased by $\delta T_{i,j}(t)$, and is denoted as:
    \begin{equation} \label{equation_II_D_2}
        T_{i,j}^{k_2}(t)=T_{i,j}^{k_2}(t) + \delta T_{i,j}^{k_1,k_2}(t)
    \end{equation}
The origin travel demand matrices for transport modes $k_1$ and $k_2$ are denoted as $[T_{i,j}^{k_1}(t)]$ and $[T_{i,j}^{k_2}(t)]$, separately, while the mode-shift adjusted matrices can be illustrated by $[T_{i,j}^{k_1, \delta}(t)]$ and $[T_{i,j}^{k_2, \delta}(t)]$, for each. We make the observation that a similar approach can be undertaken regardless of the number of modes in the network.

\subsubsection{Loyal and flexible travelling to route choice modelling}\label{II_D_2_loyal_flexible}
When disruptions occur, the change of travel cost and link properties will influence the shortest path searching at the process of trip assignment (see \textit{Step 6} in \cref{Fig_research_framework}). This logic captures the reality that travellers value travel costs before or during their travelling, which is reflected by the frequency of decision-making on mode and route. Therefore, the travellers who have loyal travel behaviour tend to travel by following their initial shortest path, which is searched before departing; they will not make changes even if disruptions occur. However, travellers who are more sensitive to travel costs tend to re-update their shortest path more frequently. The shortest path searching frequency is denoted by $\lambda$, and for those travellers that are loyal to their daily route, $\lambda$ equals the simulation period. This is means that the shortest path is only calculated once during the simulation period. For flexible travellers, the $\lambda$ is set as 10 minutes, which means that the shortest path is re-adjusted every 10 minutes according to the updated traffic states. This concept is applied in \textit{Scenario 3} and further detailed in \cref{III_D_Investigating_impact_of_mode_route_shift}. 
This decision-making of mode and route can be simulated by a discrete choice model such as the multinomial logit model (ML) through a function of utilities of all path alternatives. The impact of a travel cost change on decision-making can also be well illustrated by applying a scale parameter to the utility in the multinomial logit model; therefore, the ML model is adopted in this research study.

\section{Case study}\label{III_Case_study} 
\subsection{Geography information} \label{III_A_1_geo_information}
The case study model is implemented in the Aimsun software and represents the city of Tarragona in Spain, which contains 15 centroids, 71 nodes and 201 links. The nodes consist of 13 signalled nodes (signalised intersections) and 58 connection nodes (unsignalised intersections or turns). The links contain 201 road sections, including primary streets, freeways, ramps, roads, roundabout streets, secondary streets, urban roads, tertiary streets and toll roads with different link capacities and pre-defined limited travel speeds. There are 15 bus lines serving this area with 38 public transport stops. The land use data along with the model is collected from the census in 2012, and the initial travel cost data is generated from the trip matrices united in the vehicle for car and public transport users.

\subsection{Hypothesised disruption details} \label{III_A_1_Hypothesised_disruption_details}
\begin{figure}[ht]
 \centering
	\includegraphics[width=0.36\textwidth, height = 7.5cm]{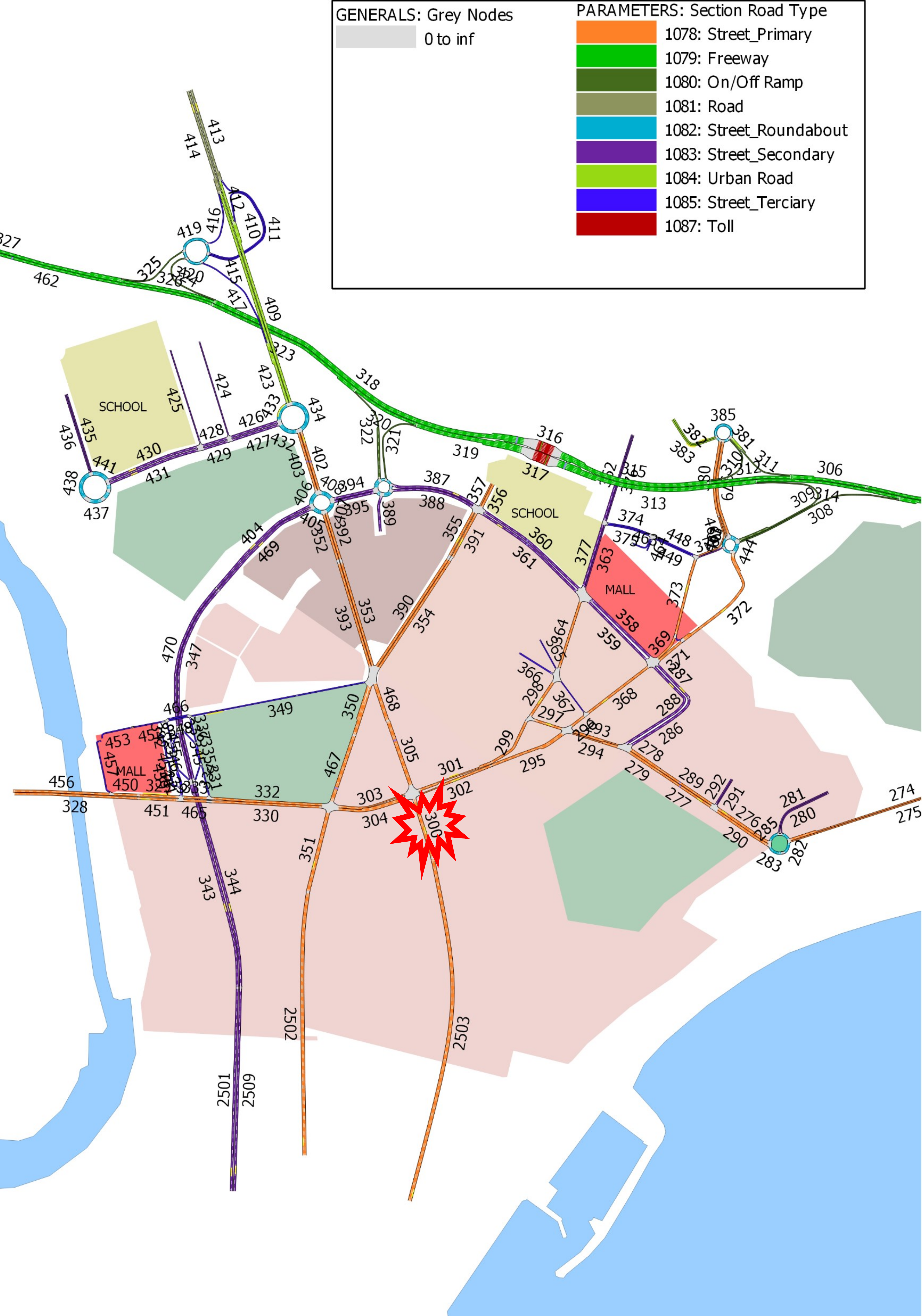}
	\caption{Map of the Tarragona area showing road networks and the disruption location.}
	\label{III_A_1_Map_Tarragona}
\end{figure}

\begin{figure*}[!htb]
    \centering
    \includegraphics[width=0.95\textwidth]{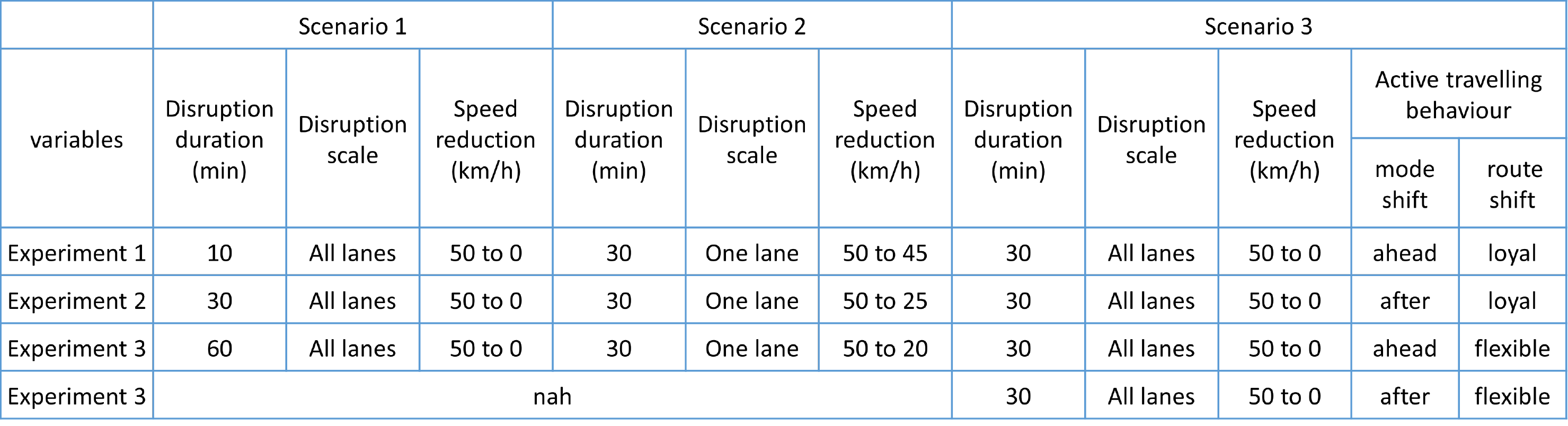}
    \caption{Details of scenario and experiments.}
    \label{III_A_1_Scenario_experiments}
\end{figure*}

To showcase our approach, we study the impact of the traffic disruption that took place at the road section 300, as highlighted by the red shape in \cref{III_A_1_Map_Tarragona}
The definition and characteristics of the modelled disruptions are further categorised and described by the following scenarios, where \textit{Scenario 1} and \textit{Scenario 2} intend to explore the consequences of the disruption in time and space, while \textit{Scenario 3} attempts to understand the impact of mode shift on the network performance as well as the network mobility.
The setting details of each scenario and experiment are presented in \cref{III_A_1_Scenario_experiments}
In \textbf{Scenario 1}, we conduct three experiments categorised by the disruption duration which can last 10, 30 or 60 minutes, according to the  \cref{equation_II_C_16} and the explanation provided in \cref{II_C_2_disruption_modelling} regarding the link closure. Any vehicle impacted by the disruption drops the travel speed to 0 km/h at the section. These experiments are named as \textit{whole lane suspended for 10, 30 and 60 minutes} in the next sections. 

\textbf{Scenario 2} contains a further three experiments defined by the disruption scale, when only one lane is impacted by the incident for 30 minutes while the rest of the lanes on the affected section remain functional, but the travel speed is reduced by 5, 25 and 30 km/h from 8:00 to 8:30 AM for each experiment. Such scenarios describe the link capacity reduction as expressed in \cref{II_C_2_disruption_modelling}. These experiments are named as \textit{single lane speed drop by 5, 25 and 30 km/h} in the next sections. 

In \textbf{Scenario 3}, we block the entire section from 8:00 to 8:30, and the travel speed drops to zero km/h during the disruption, where the mode choice ahead or after represents the situation when travellers are notified of the disruption before departing or after being blocked by a disruption; the loyal or flexible route choice is represented by the frequency of the shortest path searching during the simulating period. There are four experiments included in this scenario: 
\begin{itemize} 
\item  \textit{Experiment 1 (S3E1):} considers the situation when travellers are notified of the disruption before departing and make the decision of their mode choice ahead of the travelling, then choose their regular transport mode for the entire simulation. The new mode choice influenced by the disruption is identified by using the ``shifted ahead'' travel demand at the beginning of the simulation (this means that the travel utility calculation for the shortest path searching only happens once at the beginning of the simulation). These experiments are named as \textit{mode shift ahead, loyal to route choice} in the following subsections; 
\item  \textit{Experiment 2 (S3E2):} expresses the circumstances that, before the disruption, travellers follow the baseline demand and travel loyally according to their normal route choice; however, after the disruption occurs, they choose a new mode choice for finishing their trips. The travel utility for the shortest path searching is calculated twice: at the beginning of the simulation and after the shifted demand is applied at 8:30 AM (after the disruption occurrence) in order to simulate the updated route choice for passengers. These experiments are further referred to as \textit{mode shift after, loyal to route choice}; 
\item  \textit{Experiment 3 (S3E3):} illustrates the same setups as for \textit{Experiment 1} but instead, we are searching for the best path every 10 minutes to simulate a flexible travel behaviour for travellers. This experiment is referred to as\textit{mode shift ahead, flexible to route choice} in the following subsections; 
\item \textit{Experiment 4 (S3E4)}: demonstrates the setups that include both a mode and a flexible route choice (meaning all travellers can switch between any type of mode and also between their routing to finish their destination). This experiment follows closely real-life behaviour and is referred to as \textit{mode shift after, flexible to route choice} in the following sections. 
\end{itemize}

For all modelled experiments in all three scenarios, 
the blockage clear time is estimated based on each vehicle following the rule that the vehicle departs immediately at the end of the blockage, which is the end time of disruption ($t^{impact-end}$). The earlier arrived vehicles depart ahead of the later blocked vehicles.

\section{Results}\label{IV_Results} 
\subsection{Scenario 1 results: impact of various disruption duration} \label{III_B_Investigating_impact_of_disruption_duration}
\begin{figure}[ht]
 \centering
	\includegraphics[width=0.45\textwidth, height=7cm]{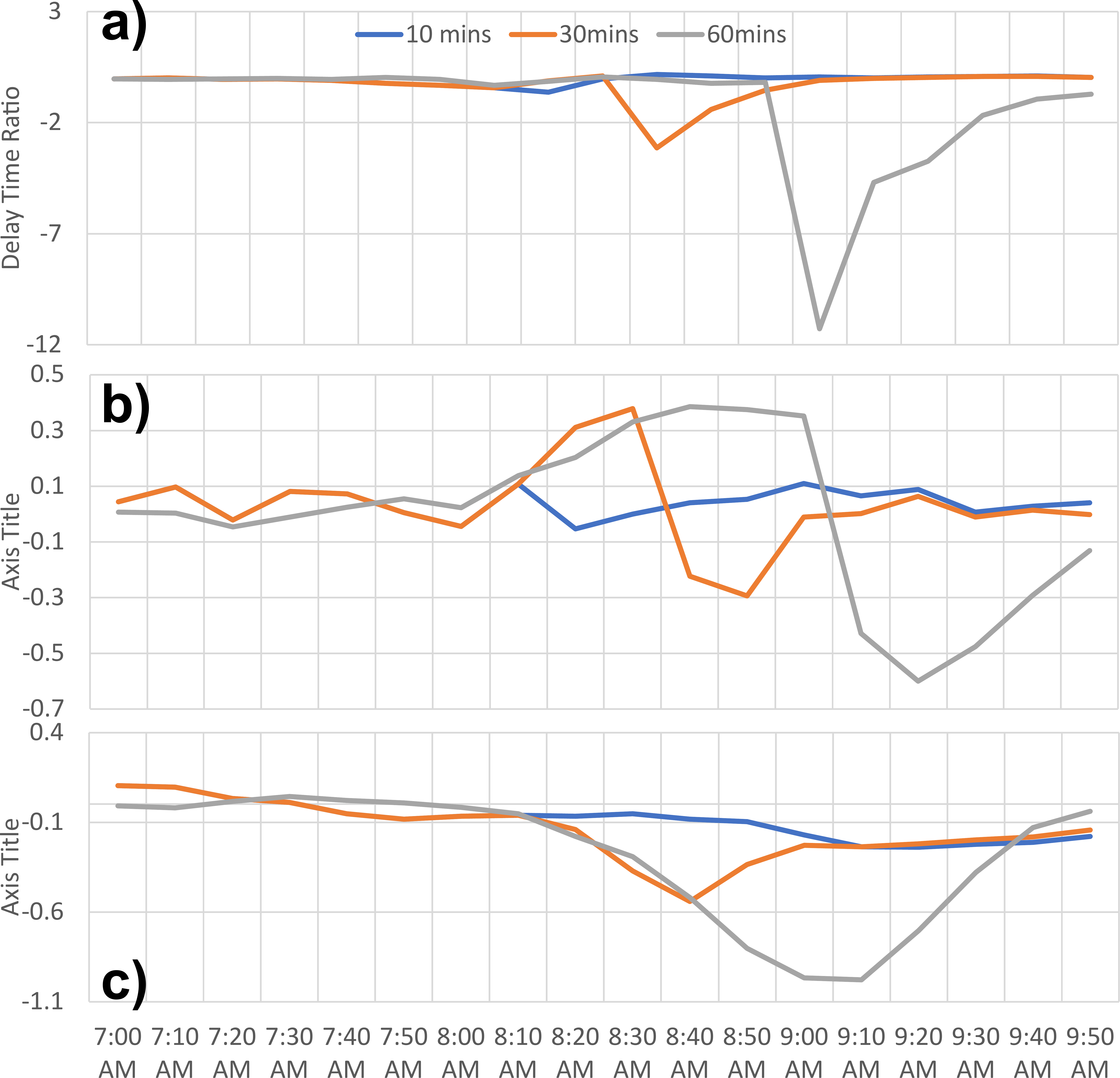}
	\caption{Impact of various disruption duration on a) delay time, b) flow and c) density }
	\label{duration_parameters}
\end{figure}

\begin{figure}[ht]
 \centering
	\includegraphics[width=0.45\textwidth, height=7cm]{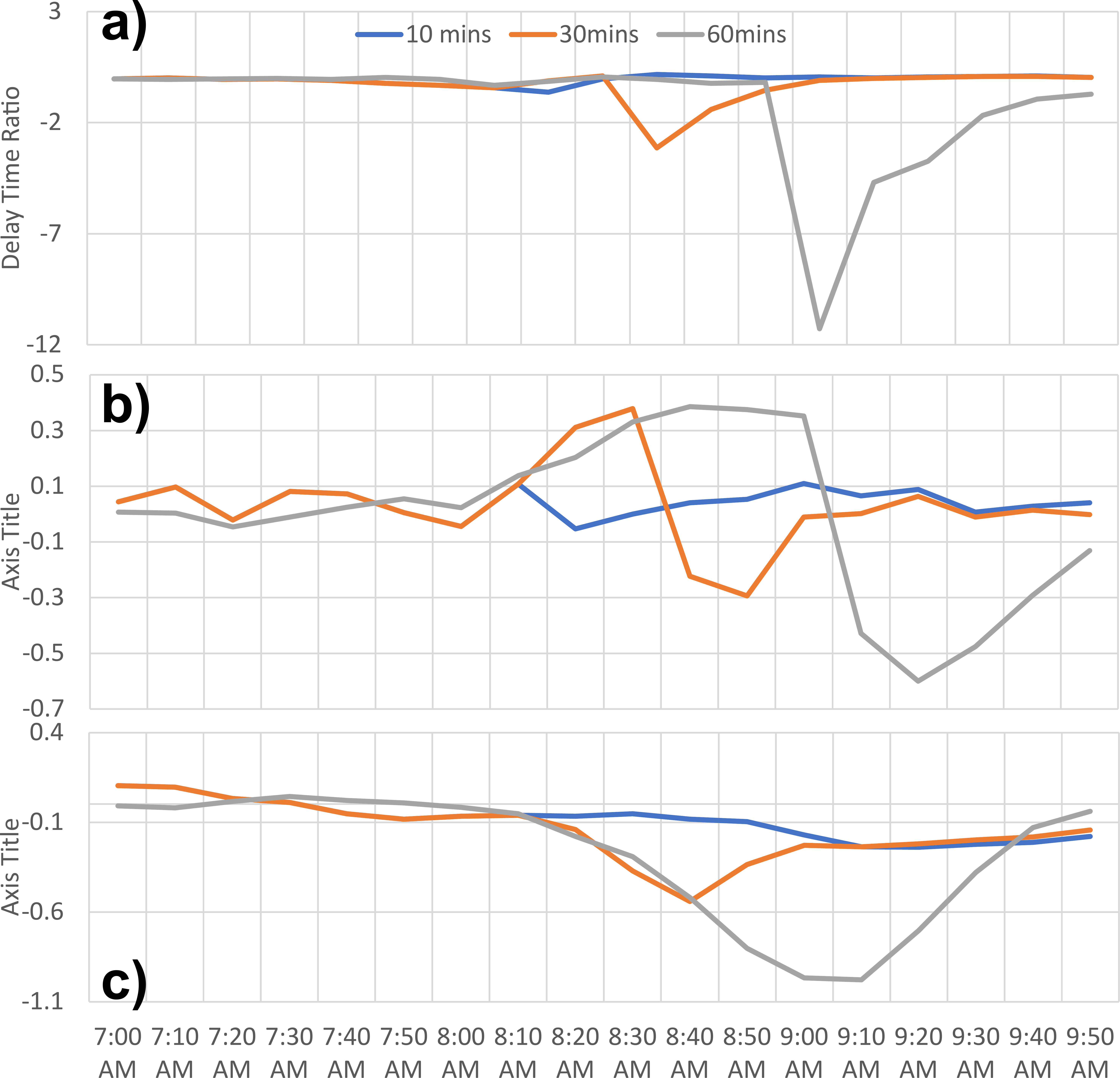}
	\caption{Impact of various disruption duration on a) delay time ratio, b) flow ratio and c) density ratio}
	\label{duration_parameters_ratio}
\end{figure}

The results of modelling the impact of various disruptions according to \textit{Scenario 1} are shown in \cref{duration_parameters} and \cref{duration_parameters_ratio}. These two figures demonstrate the time-dependent change of traffic states, namely delay time, flow and density, using a link closure as the disruption modelling method (defined in \cref{II_C_2_disruption_modelling}). The traffic states ratios shown in \cref{duration_parameters_ratio} are calculated by using \cref{equation_II_C_1}. The results are compared to the \textit{Baseline} states, which allows us to investigate the impacts of various disruptions from the baseline.

\cref{duration_parameters} and \cref{duration_parameters_ratio} combines the outputs for the morning peak hours, where the disruptions occur at 8:00 AM and last for 10, 30 or 60 mins. The mean delay time shown in \cref{duration_parameters}a) shows a significant impact post-accident, especially after 08:30-9 AM. When comparing the delay time of the baseline and their impact ratios (\cref{duration_parameters}a and \cref{duration_parameters_ratio}a) we observe that: a) for a 10-min disruption ending at 08:20 AM, there is a delay time ratio of -63\%, b) for a 30-min disruption occurring at 08:40, the delay ration quickly reaches -312\% (4.59 times higher than that of a small incident); and this is more severe c) for a 60-min disruption reaching its delay peak at 09:10 AM, when the delay ration is -1128 (more than ten times from the baseline).

\subsection{Scenario 2 results: impact of various disruption scales} \label{III_C_Investigating_impact_of_disruption_scales}
\begin{figure}[ht]
 \centering
	\includegraphics[width=0.45\textwidth, height=7cm]{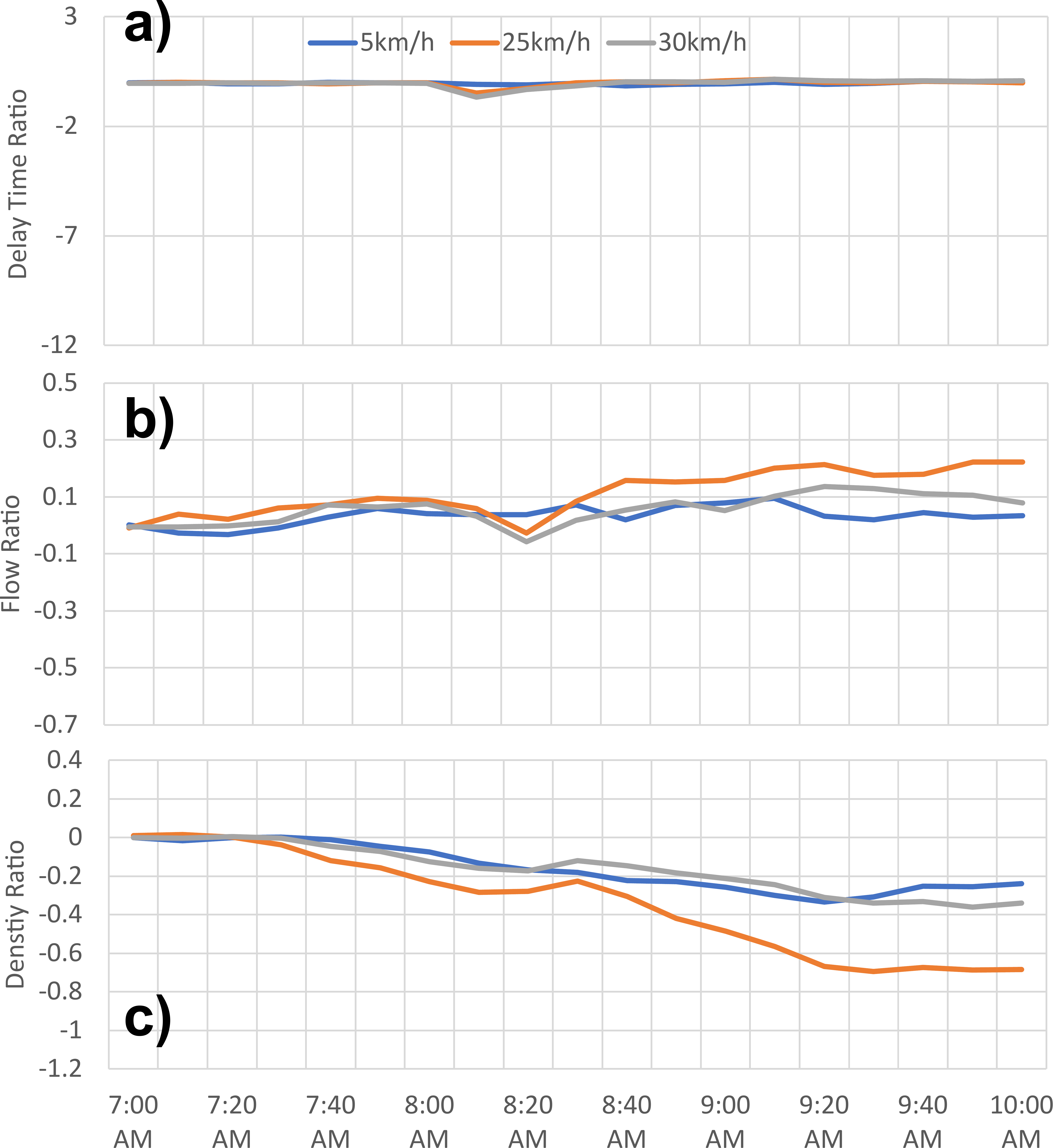}
	\caption{Impact of various disruption scales on a) delay time ratio, b) flow ratio and c) density ratio}
	\label{scale_parameters}
\end{figure}

The temporal impact of a disruption is also highly related to the disruption speed scale, as shown in \cref{scale_parameters}, where the curves of the delay time, flow and density ratios are presented. Following the \textit{Scenario 2} details from \cref{III_A_1_Hypothesised_disruption_details} and the disruption modelling method externalised by a capacity reduction in \cref{II_C_2_disruption_modelling}, the result indicates that the severity of the disruption against delay time increases with the single lane speed drop. From \cref{scale_parameters}a), we can see that the delay time does not change much in the three experiments, though there is still a peak in delay time ratio which appears at the same time (around 8:10 AM). This means that the scales of disruption could hard influence the occurrence of the peak delay. \cref{scale_parameters}b) and \cref{scale_parameters}c) show that the hypothesised single-lane speed drop also has a slight negative impact on the traffic flow and density. As shown by curves, a speed drop by 30 km/h (orange curve) results in more change in flow but it rarely impacts the change in density.

\subsection{Scenario 3 results: impact of mode and route shift} \label{III_D_Investigating_impact_of_mode_route_shift}
The impact of mode shift is analysed according to the \textit{Scenario 3} defined in \cref{III_A_1_Hypothesised_disruption_details}, where the main strategy is a travel demand adjustment: affected passengers switch from the impacted public transport to cars following either a loyal and/or flexible route choice behaviour. At this stage, for those experiments applying mode shift ahead of departing (\textit{S3E1, S3E3}), the adjusted public transport demand is increased while the car demand is decreased from the beginning of the simulation. While for the experiments that apply the mode shift after disruption (\textit{S3E2, S3E4}), the demand adjustments are conducted after the disruption ends. The loyal and flexible route choice behaviours have been depicted in \cref{II_D_2_loyal_flexible}. 

\begin{figure}[ht]
 \centering
	\includegraphics[width=0.45\textwidth]{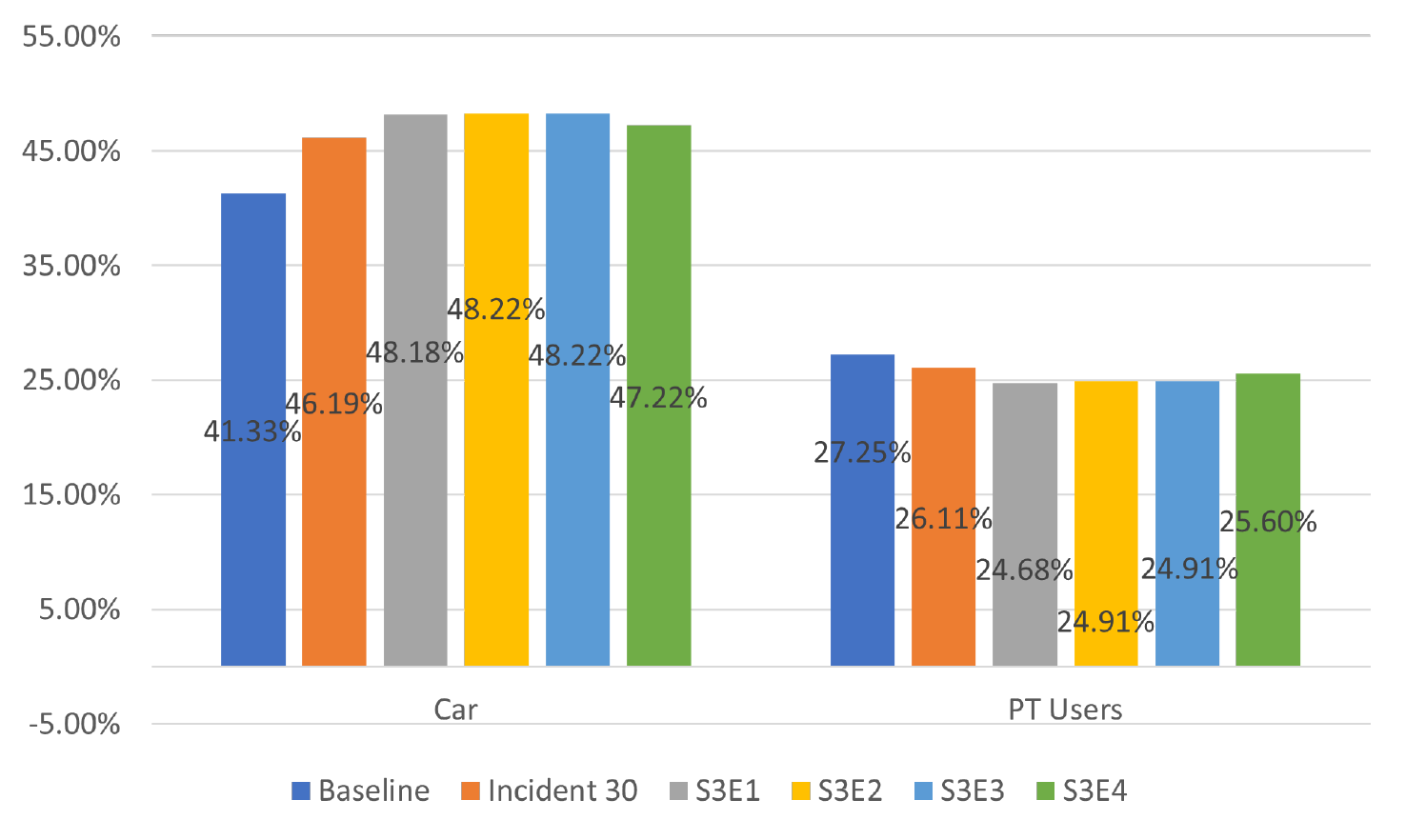}
	\caption{Comparison of travel distribution by transport modes}
	\label{travel_distribution_by_modes}
\end{figure}

The shifting on travel demand (\textit{Step 7}) is illustrated in \cref{Fig_research_framework} and after another round of dynamic trip assignments based on the shifted dynamic travel demands via the four scenarios (\textit{Step 8-11}), we analyse the \textit{Outputs 8-11} which reflect the impact of mode shift and route shift on the road network performances, as shown in \cref{travel_distribution_by_modes} and \cref{impact_mode_route_choice}. 

\cref{travel_distribution_by_modes} provides a comparison of the estimated travel demand across morning peak hours (7:00-10:00 AM), for the following experiments: \textit{Scenario 1 Experiment 2 (S1E2: whole lane suspended for 30 minutes)}, \textit{Scenario 3 Experiment 1 (S3E1: mode shift ahead, loyal to route choice)}, \textit{Scenario 3 Experiment 2 (S3E2: mode shift after, loyal to route choice)}, \textit{Scenario 3 Experiment 3 (S3E3: mode shift ahead, flexible to route choice)} and \textit{Scenario 3 Experiment 4 (S3E4: mode shift after, flexible to route choice)}.

As shown in \cref{travel_distribution_by_modes}, the mode share maintains the same for experiments \textit{Baseline} and \textit{S1E2}, while after the demand adjustment ahead of travelling (see results of \textit{S3E1} and \textit{S3E3}), the shares of car increase more dramatically than the demand adjustment after disruption (see results of \textit{S3E2} and \textit{S3E4}) when the demands for public transport decrease contrarily. The impacts of mode and route shift on traffic states are further discussed in following \cref{III_D_2_Impact_of_mode_shift} and \cref{III_D_3_Impact_of_route_shift}. 

\subsubsection{Impact of mode shift}\label{III_D_2_Impact_of_mode_shift}
\begin{figure}[h]
 \centering
	\includegraphics[width=0.45\textwidth, height=7cm]{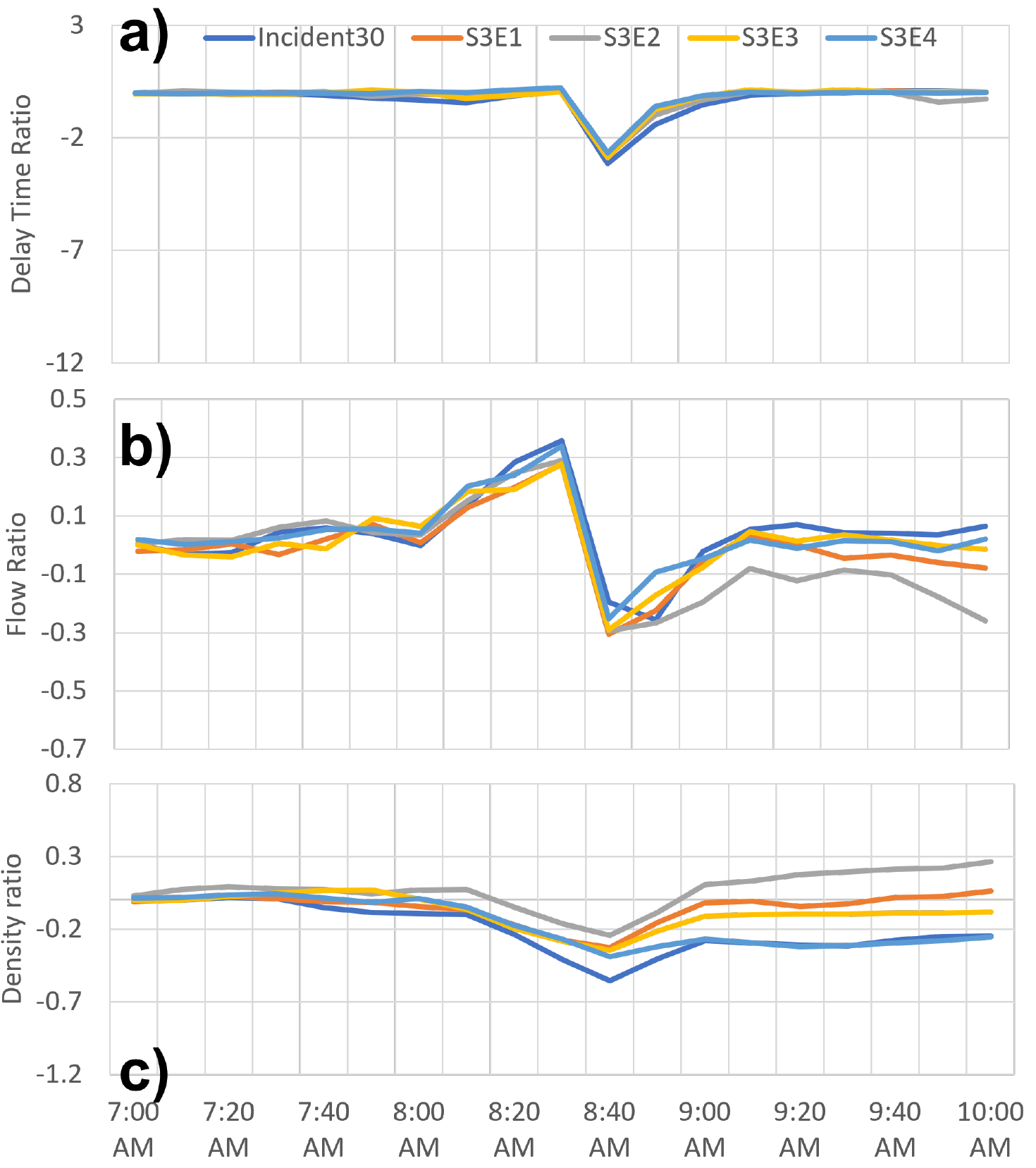}
	\caption{Comparison of the impact of mode and route choice on a) delay time ratio, b) flow ratio and c) density ratio}
	\label{impact_mode_route_choice}
\end{figure}

The impact of mode shift is compared by analysing the delay time ratio $R((t_l^{delay} (t))$, the flow ratio $R(f_l^{flow} (t))$ and the density ratio $R(f_l^{density} (t))$, based on experiments \textit{S3E1} and \textit{S3E2}. As shown in \cref{impact_mode_route_choice} a), both modes shift ahead and after the disruption benefit the travellers by reducing delay time, and the mode shift ahead (\textit{S3E4}) is slightly superior to mode shift after (\textit{S3E2}) as can be observed by analysing the timing around 8:50 to 9:30 in this case.
By looking at the curves of flow ratio (\cref{impact_mode_route_choice} b)) and density ratio (\cref{impact_mode_route_choice} c)), both mode shift strategies increase travel flow after disruption and ease the traffic congestion. The curves of flow ratio indicate that before and during the disruption, the mode shift ahead increases the flow, but due to the extra demand, the performance of the post-disruption congestion relief is limited. An interesting phenomenon is that, right after the disruption, the road section is free of the vehicle; this enables the initially released vehicles to run in a free-flow situation, which results in a peak of flow that is even higher than the \textit{Baseline} situation. 
The density curves in \cref{impact_mode_route_choice}c) indicate that shifting modes after the disruptions is more efficient on easing the congestion because the bottom density ratio is reduced only to $-24\%$ for \textit{S3E2} (see grey curve) while for \textit{S3E1} it can only touch down to $-33\%$, and the bottom density ratio is $-34\%$ for \textit{S3E3} (yellow curve) while for \textit{S3E4} it is $-39\%$, comparing with the ratio of \textit{Incident 30} (dark blue curve) is $-56\%$. 
\subsubsection{Impact of route shift}\label{III_D_3_Impact_of_route_shift} 
The route shift is modelled using a stochastic route choice, where trips can be assigned to the network after the shortest travel path for each trip, as introduced in \cref{II_D_2_loyal_flexible}. To understand the impact of route shift on network performance, results based on \textit{S3E1} verses \textit{S3E3}, and \textit{S3E2} verses \textit{S3E4} are selected accordingly. The significant benefit in delay time reduction can be observed when applying the \textit{S3E4} with mode shift applied after disruptions and flexible route choice behaviour, where the delay time ratio is reduced to $-267\%$ and the ratio of flow is reduced to $-25\%$ (see \cref{impact_mode_route_choice} a) and b)). According to \cref{impact_mode_route_choice} c), the higher density ratio appears on the curve of \textit{S3E2}($-24\%$), which indicates that the strategy with flexible route choice with mode shift ahead performs much more gratifying. 

\section{Conclusion}\label{V_Conclusion} 
This paper proposed an integrated multi-modal hybrid modelling framework that embeds the four-step model estimation with a dynamic demand estimation and mode shift approach. This framework demonstrates the potential ability to model real-time disruptions, their impact and an evaluation of mode shift and route shift in a multi-modal environment. We consider the essential of dynamical microscopic transport simulation and propose a methodology to identify and quantify the temporal-spatial impact of transport disruption. For the spatial impact analysis, we proposed a method to detect the impacted trips between OD pairs. This permits us to simulate the mode choice behaviour without starting from the static demand estimation, which, therefore, does not require more land-use data. The results and findings generated from this research study evidence that public transport does make travel more accessible, especially when disruption is suspending the traffic. The mode and route shift benefit the transport system by increasing the flow and effectively reducing delay time and density. 
Future investigations could cover the topic of examining more types of disruptions by duration or location by links. The theoretical link-based disruption impact estimation method is proposed in this paper, but the paper falls under statistical analysis following this idea. More investigation towards impact propagation into the network through links could be a good sub-direction. The impact of dual or multiple disruptions on network performance is also underestimated in this paper. We limit the work only by considering the single disruption, single land closure and single road section closure. The method to quantify the impact of multiple disruptions in the network is also required. Investigating data-driven methods for impact measurement and model calibration is another highly recommended direction.
\section*{Acknowledgments}
This research is supported by the ARC LP project LP180100114.


%
%


\bibliographystyle{IEEEtran}
\bibliography{UTS_PhD-IEEE_ITSC_2021}



\end{document}